\documentclass[aip,jcp,reprint,eqsecnum,a4paper]{revtex4-1}

\usepackage{graphicx}
\usepackage[sort&compress]{natbib}
\usepackage[sumlimits]{amsmath}
\usepackage{braket}
\usepackage{amsfonts}

\usepackage{amssymb}
\usepackage{dcolumn}
\usepackage{subdepth}
\usepackage{color}
\usepackage{enumitem}
\usepackage{blindtext}
\usepackage{bm}

\usepackage[colorlinks=true,linkcolor=red,citecolor=blue, bookmarksopen=true]{hyperref}

\begin{document}

\title{Shaping the Laser Control Landscape of a Hydrogen Transfer Reaction by Vibrational Strong Coupling. A Direct Optimal Control Approach}

\author{A. R. Ramos Ramos}
\affiliation{Institute of Physics, University of Rostock,
  Albert-Einstein-Stra{\ss}e 23-24, D-18059 Rostock, Germany}
\author{E. W. Fischer}  
\affiliation{Institut f\"ur Chemie, Humboldt-Universit\"at zu Berlin, Brook-Taylor-Stra{\ss}e 2, D-12489, Berlin, Germany}
\affiliation{Institute of Chemistry, University of Potsdam, Karl-Liebknecht-Stra\ss{}e 24-25, D-14476 Potsdam-Golm, Germany}

\author{P. Saalfrank}  
\affiliation{Institute of Chemistry, University of Potsdam, Karl-Liebknecht-Stra\ss{}e 24-25, D-14476 Potsdam-Golm, Germany}

\author{O. K\"uhn}
\affiliation{Institute of Physics, University of Rostock,
  Albert-Einstein-Stra{\ss}e 23-24, D-18059 Rostock, Germany}

\begin{abstract}
Controlling molecular reactivity by shaped laser pulses is a long-standing goal in chemistry. Here we suggest a direct optimal control approach which combines external pulse  optimization with other control parameters arising in the upcoming field of vibro-polaritonic chemistry, for enhanced controllability
The direct optimal control approach is characterized by a simultaneous simulation and optimization paradigm, meaning that the equations of motion are discretized and converted into a set of holonomic constraints for a nonlinear optimization problem given by the control functional. Compared with indirect optimal control this procedure offers great flexibility such as final time or Hamiltonian parameter  optimization. Simultaneous direct optimal control (SimDOC) theory will be applied to a model system describing H-atom transfer in a lossy Fabry-P\'erot cavity under vibrational strong coupling conditions. Specifically, optimization of the cavity coupling strength and thus of the control landscape will be demonstrated.
\end{abstract}

\maketitle


\section{Introduction}
\label{sec:intro}

The emerging field of vibro--polaritonic chemistry studies the impact of light--matter hybrid states known as vibrational polaritons on chemical reactivity.\cite{hutchison_modifying_2012,hirai_2020,nagarajan_2021,dunkelberger_2022} Vibrational polaritons are formed once the interaction between molecular vibrations and quantized light modes of optical Fabry--P\'erot cavities reaches the vibrational strong coupling (VSC) regime \cite{ebbesen_2016} and are experimentally well characterized by different spectroscopic approaches.\cite{shalabney_2015,george_2016,xiang_2018} At the heart of vibro--polaritonic chemistry lies the goal of controlling thermal chemical reactions by means of cavity (vacuum) fields, whose coupling to the molecular reaction coordinate modifies the potential energy landscape.\cite{nagarajan_2021,simpkins_2023}
The possibility of altering chemical reactions by means of quantized fields of optical cavities has been reported in a series of seminal experiments, where cavity modes were tuned resonant to either reactant or solvent modes.\cite{george_2015,thomas_2016,lather_2019,thomas_2019}

The use of classical laser fields to control chemical reactions has been extensively explored since the 1980s (for an early review, see Ref. \citenum{manz97_80}, a more recent overview is given in Ref. \citenum{keefer18_2279}). Focusing on the infrared (IR) laser--driven dynamics in the electronic ground state, the number of experimental realizations is limited. Vibrational excitation via ladder climbing has been demonstrated for several small molecules,\cite{arrivo95_247,maas98_75,strasfeld07_038102} even triggering bond dissociation.\cite{witte03_2021} A prominent example is the isomerization reaction of HONO, demonstrated in Ref. \citenum{schanz05_044509}. Only recently it has been possible to control bimolecular reactions in solution with IR light.\cite{stensitzki18_126,heyne19_11730} Intramolecular vibrational energy redistribution (IVR) and vibrational energy relaxation (VER) into the solvent are typically considered a limitation of IR laser control of vibrational dynamics. On the other side, IVR, i.e., anharmonic couplings, can be an essential part of the control strategy in cases where the reaction coordinate does not have a large overlap with IR--active modes.\cite{windhorn03_641} The hybridization between molecular vibrational states and photon states of the cavity potentially can change IVR and VER pathways.\cite{simpkins_2023}

Given these developments, it seems natural to combine vibro--polaritonic chemistry with laser control by means of an external field.
While laser control typically works in the weak field limit, where the molecular Hamiltonian is given, VSC provides a means to modify this Hamiltonian thus changing the control landscape. 
Indeed recently, a combination of both  control theory and strong coupling in optical cavities has been theoretically proposed  for a generic two level model.\cite{flick_2017} Notably, different strategies have been explored in the literature, employing either quantum \cite{carnio_optimization_2021,lindel_2023} or classical\cite{flick_2017,bergholm_optimal_2019,fan_quantum_2023} external fields. 

An early example for theoretical IR laser control of reactions is the H--transfer in thioacetylacetone (TAA), where it was shown that laser fields can be used to trigger quantum tunneling through the reaction barrier. \cite{doslic_1999,doslic00_247,kuhn00_6104,doslic11_411} Recently, this reaction was revisited in Ref. \citenum{fischer_saalfrank_2023} where the isomerization has been modeled as a population transfer in a cavity--distorted double--minimum reaction potential without an external field. Results from this paper showed that transfer rates from the enol to the enethiol configuration significantly increase with light--matter interaction strength. 
This is a direct consequence of the modification of the reaction potential in the VSC regime. Indeed the effect of VSC on the reaction is more conveniently viewed in terms of vibro--polaritonic hybrid states on the global cavity potential energy surface (cPES) including molecular and quantum field coordinates.
Given these two separate studies, the present contribution sets its focus on the laser control of H--atom transfer on the combined vibro--polaritonic cPES, thus exploring the possibilities provided by shaping the control landscape via VSC.

Optimal control theory (OCT) provides a rigorous variational tool for the design of external laser fields subject to certain constraints. In Molecular Physics, OCT is mostly used in its \textit{indirect} formulation. It starts with an analytical derivation of the stationarity condition of the control functional. Subsequently, numerical procedures need to be derived and implemented to solve the resulting equations, e.g. Ref. \citenum{werschnik07_R175}. Within this category fall the Krotov \cite{reich_monotonically_2012} and the GRAPE (gradient ascent pulse engineering) \cite{khaneja_optimal_2005} methods. In contrast \textit{direct} OCT follows the first--discretize--then--optimize paradigm. Here the control trajectories are parameterized in time to transform the performance functional to be optimized into a performance function depending on the parameters of the control.\cite{kelly_introduction_2017} There are two variants: First, a sequential solution (SeqDOC -- Sequential Direct Optimal Control), where at each optimization step the dynamic equations of the system are solved to evaluate the performance function. Most notable examples are the GOAT  (gradient optimization of analytic controls)\cite{machnes_tunable_2018} and the DCRAB (dressed chopped random basis) methods. \cite{muller_one_2022} In the second variant also the state trajectories are discretized in time, transforming the equations of motions into a set of holonomic constraints for the performance function (SimDOC -- Simultaneous Direct Optimal Control). This variant is more common in engineering applications.\cite{kelly_introduction_2017} Its main advantage over both indirect and sequential direct OCT is the flexibility by which constraints can be incorporated, i.e. they are part of the discretization and do not require, e.g., analytical calculations or algorithmic developments. In the context of Molecular Physics, applications of SimDOC have been presented in Refs. \citenum{ramos_ramos_direct_2021,ramos_ramos_manipulating_2022}.  Here we will extend our previous investigations of direct OCT  to a novel  application, i.e. laser control of vibro--polaritonic chemistry. Further, we will demonstrate the strength of the method by showing that it is possible to optimize parameters of the Hamiltonian itself (the magnitude of VSC) and the final time entering the control functional.

The paper is organized as follows: In section \ref{sec:model} the model Hamiltonian will be defined and section \ref{sec:opt_scheme} will give an overall description of the optimization algorithm. After specification of the model parameters in section \ref{sec:model_par},  sections \ref{sec:tf_opt} and \ref{sec:lm_inter_opt} will present, respectively, the results for the final time and cavity--matter interaction strength optimization. Section \ref{sec:omega_c} deals with the influence of the cavity frequency on the control strategy, and in section \ref{sec:conlcusions} the conclusions will be summarized.

\section{Theoretical Model}

\subsection{Model Hamiltonian}\label{sec:model}
We apply the SimDOC approach to the  asymmetric hydrogen transfer reaction model of TAA \cite{doslic_1999} interacting with a single quantized infrared cavity mode. Reaction coordinate and cavity mode will be denoted by $q$ and $x_{\rm c}$, respectively. The full time--dependent Hamiltonian, $\hat{H}(t)$, is composed of a light--matter hybrid contribution, $\hat{H}_{0}$, which is driven by an external classical laser field via $H_{\rm F}(t)$, and reads in the Schr\"odinger picture
\begin{equation}
	\hat{H}(t)=\hat{H}_{0} +H_{\rm F}(t)~,
\end{equation}
with the driving term in dipole approximation being
\begin{equation}
    H_{\rm F}(t)=-\mu(q) E(t)~.
\end{equation}
Here, $\mu(q)$ is the molecular dipole moment and $ E(t)$ is the external electric field. Note that we omit here the vector character of both dipole moment and field, and we assume that the dipole moment depends on the reaction coordinate only.
For the system Hamiltonian, $\hat{H}_{0}$, we consider an effective vibrational Pauli--Fierz Hamiltonian in length--gauge representation, crude cavity Born--Oppenheimer and long--wavelength approximation \cite{flick_2017cbo,schaefer_2018,fischer_ground_2021,fischer_saalfrank_2023ccbo}
\begin{equation}
\label{eq:htot}
	\hat{H}_0=\hat{H}_{\rm S}+\hat{H}_{\rm C}+V_{\rm SC}+V_{\rm DSE}\, ,
\end{equation}
initially studied for the TAA--cavity hybrid system in Ref. \citenum{fischer_saalfrank_2023}. The first term constitutes a one--dimensional reaction Hamiltonian (we set $\hbar=1$)
\begin{equation}
	\hat{H}_{\rm S}=-\frac{1}{2 \mu_S} \frac{\partial^2}{\partial q^2} + V(q) \, ,
	\label{eq.model_transfer_hamiltonian}
\end{equation}
with mass $\mu_{\rm S}=1914.028\,m_{\rm e}$ and reaction coordinate potential \cite{doslic_1999}
\begin{equation}
    V(q) = \frac{1}{2}\left(V_+(q)-\sqrt{V^2_-(q)+4\,K^2(q)}\right)\,
\end{equation}
where $V_\pm(q)=V_1(q)\pm V_2(q)$, and harmonic potentials  $V_i(q)=\frac{1}{2}m_i\,\omega^2_i(q-q_{i,0})^2+\Delta_i$. Here, $V_1(q)$ corresponds to the O-H$\cdots$S configuration  and $V_2(q)$ to the O$\cdots$H-S configuration of TAA, respectively, with corresponding parameters given in Tab. \ref{tab.parameters_syspot}.
\begin{table}[hbt!]
\centering
\begin{tabular}{c c c c c}
\hline\hline
$m_1/m_{\rm e}$ & $\omega_1/ E_{\rm h}$ & $q_{1,0}/a_0$ & $\Delta_1/E_{\rm h}$
\vspace{0.2cm}\\
$1728.46$   & $0.01487$      & $-0.7181$     & $0.0$
\vspace{0.1cm}\\
\hline
$m_2/m_{\rm e}$ & $\omega_2/E_{\rm h}$ & $q_{2,0}/a_0$ & $\Delta_2/E_{\rm h}$
\vspace{0.2cm}\\
$1781.32$   & $0.01247$      & $1.2094$ & $0.003583$
\vspace{0.1cm}
\\
\hline\hline
\end{tabular}
\caption{Reaction potential parameters with upper row corresponding to $V_1(q)$ (O-H$\cdots$S) and lower row to $V_2(q)$ (O$\cdots$H-S).}
\label{tab.parameters_syspot}
\end{table}
The coupling function between the two harmonic potentials is given by $K(q)=k\,\exp(-(q-q_0))^2$ with $k=0.15582\,E_{\rm h}$ and $q_0=0.2872\,a_0$. 
Note that in Ref. \citenum{doslic_1999} a dissipative model has been proposed, where the  reaction coordinate was coupled to a primary vibrational mode corresponding to heavy atom motion, O$\cdots$S, of the hydrogen bond as well as to a not further specified heat bath. These contributions are not considered in the present model, which has the focus on the interaction with the cavity.

The single--mode cavity Hamiltonian in Eq.~\eqref{eq:htot} in coordinate representation is given by
\begin{equation}
	\hat{H}_{\rm C}=-\frac{1}{2}\frac{\partial^2}{\partial x_{\rm c}^2} + \frac{\omega_{\rm c}^2}{2}x_{\rm c}^2 \, ,
\end{equation}
with cavity displacement coordinate, $x_{\rm c}$, and harmonic cavity frequency, $\omega_{\rm c}$. Further, the cavity--matter interaction potential reads
\begin{equation}
	V_{\rm SC}=\sqrt{2\omega_{\rm c}}\, g\, \mu(q)\, x_{\rm c} \, ,
\end{equation}
with cavity--matter interaction constant $g$, which has dimensions of an electrical field strength. Here, we express $g$ in terms of a dimensionless interaction parameter $\eta$ via $g=\frac{\omega_c}{\mu_{10}^q}\eta$, where $\mu_{10}^q$ is the fundamental vibrational transition dipole moment of the one--dimensional reaction model Hamiltonian, Eq.~\eqref{eq.model_transfer_hamiltonian}. Eventually, the dipole self--energy is given as
\begin{equation}
	V_{\rm DSE}=\frac{g^2}{\omega_{\rm c}} \mu(q)^2 \, .
\end{equation}
For the molecular dipole function, we employ a linear approximation \cite{doslic_1999}
\begin{equation}
	\mu(q)=\mu_0+\mu_1(q-q_{\rm d})\, ,
\end{equation}
with parameters, $\mu_0=1.68\,ea_0$, $\mu_1=-0.129\,e$ and $q_{\rm d}=-0.59\,a_0$. Note that within this model we neglect a possible direct excitation of the cavity mode by the external field.

In passing we note that the form of the vibrational Pauli-Fierz Hamiltonian,  $\hat{H}_{0}$, actually corresponds to a special realization of the  All-Cartesian reaction surface Hamiltonian widely used for the description of H transfer coupled to heavy atom modes.\cite{ruf_new_1988,giese_multidimensional_2006} 

Eigenstates of the effective vibrational Pauli--Fierz Hamiltonian, $\psi_m(q,x_{\rm c})=\langle q,x_{\rm c}|m\rangle$, denoted as vibrational polaritons or vibro--polaritonic states, satisfy the two-dimensional time--independent Schr\"odinger equation
\begin{equation}
    \hat{H}_{0}\psi_m(q,x_c)=E_m\,\psi_m(q,x_c)\, ,
\end{equation}
with eigenenergies, $E_m$. 

As a consequence of imperfections in the cavity mirrors, cavity mode excitations will have a finite lifetime due to spontaneous decay. This effect will be empirically included by assigning state--dependent effective decay rates
\begin{equation}
\varGamma_{m}=
2 \gamma \omega_c \sum_{n<m}\vert\braket{n\vert x_c\vert m}\vert^2
\quad,
\label{eq.state_decay_rate}
\end{equation}
which  account for the total decay amplitude of the $m^\mathrm{th}$ vibro--polaritonic state to all states lower in energy. $\varGamma_{m}$ is based on the Fermi's Golden Rule, assuming a weak coupling between the vibro-polaritonic system and the out-of-cavity radiation field, which is mediated by the cavity mode. This coupling is taken to be linear in the  cavity mode coordinate. The constant $\gamma$ is the effective decay rate of the bare cavity and can be given in terms of the experimental cavity quality factor $Q$ by $\gamma=\omega_{\rm c}/Q$.\cite{shalabney_2015}

We now consider the spectral representation of $\hat{H}(t)$ in the basis of vibro-polaritonic states with matrix elements
\begin{equation}
   H_{mn}(t) = \left(E_m-\mathrm{i}\dfrac{\varGamma_{m}}{2}\right)\,\delta_{mn}-\mu_{mn}\,E(t)\, ,
    \label{eq.matelem}
\end{equation}
where $\mu_{mn}=\langle m | \mu | n \rangle$ is a vibro-polaritonic (transition) dipole moment matrix element. Further, we augmented the diagonal by an imaginary contribution, which accounts for a finite lifetime of cavity mode excitations via state-dependent effective decay rates, $\varGamma_{m}$, defined in Eq.~\eqref{eq.state_decay_rate} in analogy to Refs. \citenum{ulusoy_vendrell_2020,fischer_saalfrank_2022vpci}. Note that according to Eq.~\eqref{eq.matelem}, the norm of the time-dependent wavefunction to be computed below is not conserved due to the imaginary decay contribution. 

The time--evolution of the laser--driven cavity--matter hybrid system is determined by the time--dependent Schr\"odinger equation in the interaction picture with respect to $\hat{H}_{0}$
\begin{equation}
\label{eq:TDSE}
i\frac{\partial}{\partial t} | \Psi^{(\rm I)}(t) \rangle = \hat{H}^{\rm (I)}_{\rm F}(t) | \Psi^{\rm(I)}(t) \rangle\, .
\end{equation}
The interaction picture yields a smoother dynamics, which is vital for efficient treatment  using direct optimal control. Eq.~\eqref{eq:TDSE} can then be projected on the basis of vibro--polaritonic eigenstates, $|m\rangle$, which gives rise to the following two equations after splitting the real $(\Re)$ and imaginary $(\Im)$ parts:
%
%
%
\begin{equation}\label{eq:re_dyn}
\begin{split}
&\frac{\partial}{\partial t}  \Re \langle m |\Psi^{(\rm I)}(t) \rangle =  \sum_n\Big[-\mu_{mn}E(t)  \Big(  \cos (\omega_{mn} t)     \\
&\Im \langle n | \Psi^{(\rm I)}(t) \rangle   +   \sin (\omega_{mn} t) \Re  \langle n | \Psi^{(\rm I)}(t) \rangle  \Big)\Big] \\
&  - \frac{\varGamma_{m}}{2}   \Re \langle m | \Psi^{(\rm I)}(t) \rangle  ~,
\end{split}
\end{equation}
\begin{equation}\label{eq:im_dyn}
\begin{split}
&\frac{\partial}{\partial t}  \Im \langle m |\Psi^{(\rm I)}(t) \rangle =  \sum_n\Big[-\mu_{mn}E(t)  \Big(  \sin (\omega_{mn} t)     \\
&\Im \langle n | \Psi^{(\rm I)}(t) \rangle   -  \cos (\omega_{mn} t) \Re  \langle n | \Psi^{(\rm I)}(t) \rangle  \Big)\Big]\\
&  - \frac{\varGamma_{m} }{2}  \Im \langle m | \Psi^{(\rm I)}(t) \rangle ~,
\end{split}
\end{equation}
where $\omega_{mn}=E_m-E_n$.

The system of dynamic equations \eqref{eq:re_dyn}--\eqref{eq:im_dyn} can be   cast to the following generic form:
\begin{equation}\label{eq:diffcon}
  \dot{\bm a}(t)=\bm f[\bm a(t),\bm u(t),\bm k,t]~,
\end{equation}
where the state vector is given by  ${\bm a}(t)=(\Re \langle m |\Psi^{(\rm I)}(t) \rangle,\Im \langle m |\Psi^{(\rm I)}(t) \rangle)^{\rm T}$, $\bm u(t)=E(t)$ is the external control (i.e. the laser field), and $\bm k$ comprises a set of static parameters.

\subsection{Optimization scheme}\label{sec:opt_scheme}

The optimal control method used in this paper is based on a SimDOC approach, see Section \ref{sec:intro}.
The discretization procedure is carried out by the open source PSOPT package \cite{becerra_solving_2010}, which utilizes the package IPOPT \cite{frank_e_curtis_note_2012} for large-scale nonlinear optimization.

The goal of the optimization can be stated as follows. Given some initial normalized state $|\Psi(t_{\rm 0})\rangle=|\Phi^{\rm i}\rangle$, characterized by the parameters $\bm a^{\rm i}$, find a laser field $E(t)$ such that the overlap is maximized between the time--evolved final state, $|\Psi(t_{\rm f})\rangle$, at $t=t_{\rm f}$ and some normalized target state $|\Phi^{\rm t}\rangle$, characterized by the parameters $\bm a^{\rm t}$. At the same time, minimize the energy input of the laser field.  To accomplish this goal, we start from a general performance functional (also called total cost below)
\begin{equation}\label{eq:J}
\begin{split}
    {\mathcal J}[\bm a, \bm u, \bm k, t_{\rm f}]= &\\
    {\mathcal T}&[\bm a(t_{\rm f}),\bm k,t_{\rm f}] +  {\mathcal R}[\bm a,\bm u,\bm k,t_{\rm f}]\,,
\end{split}
\end{equation}
that should be minimized subject to the differential constraints, Eq.  \eqref{eq:diffcon}, between $t_0$ and $t_{\rm f}$. The terminal cost is given by (notice the minus sign because the performance functional will be minimized and we want to maximize the overlap)
\begin{equation}
\begin{split}
	{\mathcal T}[\bm a(t_{\rm f}),\bm k,t_{\rm f}] & = -\left| \langle  \Phi^{\rm t}  |  \Psi(t_{\rm f}) \rangle \right|^2 \\
  & = -\left| \langle  \Phi^{\rm t}   |  e^{-i\hat{H}_{0}t_{\rm f}}  \Psi^{\rm (I)}(t_{\rm f}) \rangle \right|^2\,.
\end{split}
\end{equation}
The running cost is chosen to depend on the external control and the final time only, i.e.
\begin{equation}\label{eq:run_cost}
\begin{split}
{\mathcal R}[E(t),t_{\rm f}] &=\kappa  \int_{t_0}^{t_{\rm f}} \frac{ |E(t)|^2}{s(t)} dt,\\
\quad s(t)&=\sin^2 \left( \frac{\pi}{t_{\rm f}-t_{\rm 0}} (t-t_{\rm 0})\right) +\epsilon \, .
\end{split}
\end{equation}
 Besides the field intensity, we have included a factor $\kappa$ scaling the penalty for high field strengths as well as a shape function, $s(t)$, which  ensures that the field increases(decreases) slowly when turned on(off).\cite{sundermann_extensions_1999} Note that $\epsilon$ is a small parameter introduced to avoid division by zero and numerical problems at times $t=t_{\rm 0}$ and $t=t_{\rm f}$. We have used $\epsilon=0.005$ throughout. Finally, the penalty parameter in Eq. \eqref{eq:run_cost} was fixed at $\kappa=4.8 \times 10^{-1} \, e^2 a_0^2/ E_h$.

In general, there could be path and event constraints.\cite{becerra_solving_2010} For the application presented below, we do not use any path constraints, but event constraints. Given the event
\begin{equation}\label{event}
\bm e[\bm a(t_0),F[E(t)]]=
  \begin{pmatrix}
    \bm a(t_0) \\
    \int_{t_0}^{t_{\rm f}} E(t) dt
  \end{pmatrix}~,
\end{equation}
upper and lower bounds will be chosen equal as follows
\begin{equation}
  \bm e_{\rm L}= \bm e_{\rm U}=
\begin{pmatrix}
  \bm a^{\rm i}\\
  0
  \end{pmatrix}~.
\end{equation}
Hence, the parameters of the initial state are fixed and not subject to optimization. Further, we enforce the zero--net--force condition by demanding that $F[E(t)]= \int_{t_0}^{t_{\rm f}} E(t) dt=0$.\cite{doslic_generalization_2006} 

From the available discretization methods in the PSOPT package, we will use the trapezoidal one with 2100 time--discretization points since it offers the best compromise between accuracy and performance. One can find more information about this discretization method in Refs. \citenum{becerra_solving_2010,kelly_introduction_2017}.
 
The actual laser field optimization starts from random initial guesses for fields and states evolution. Unless stated otherwise,  a set of 10 calculations with different initial guesses has been performed for every set of parameters chosen. From these calculations, we have kept the one with smallest total cost. This procedure ensures a larger exploration of the control space and better optimal solutions.

\subsection{Model Parameters}
\label{sec:model_par}
\begin{figure}[htbp]
\centering
\includegraphics[width=\linewidth]{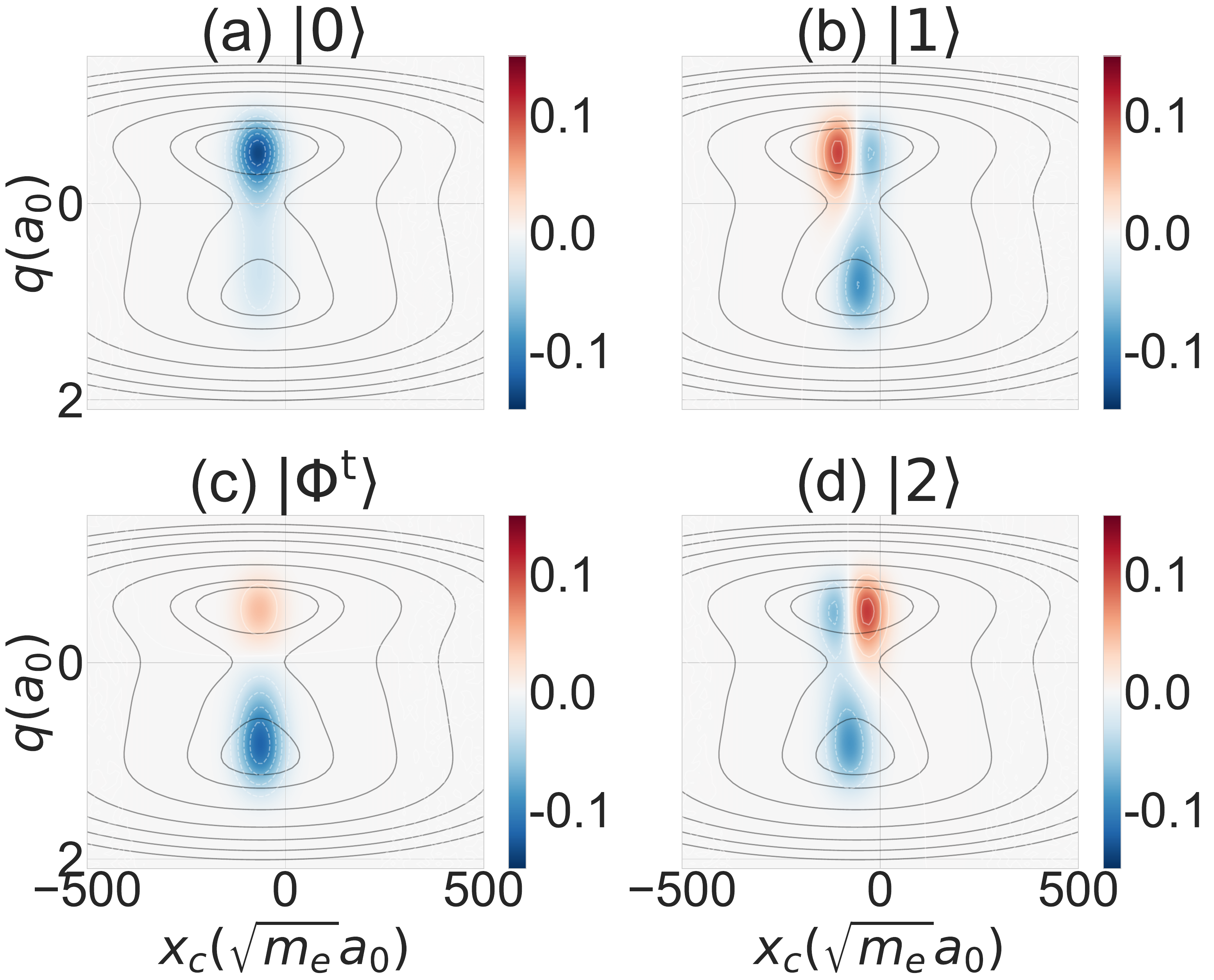}
  \caption{Vibro-polaritonic eigenstates and target state, $|\Phi^{\rm t}\rangle$ in (c), with contour plot of the two-dimensional cPES for $\eta=0.03$ and $\omega_{\rm c}=\omega_{10}^q$. The corresponding energies of the depicted states are: (a) $E_0=4.40\times 10^{-3} E_h \,(966.4\,{\rm cm}^{-1})$, (b) $E_1=4.96\times 10^{-3} E_h \,(1089.1\,{\rm cm}^{-1})$, (c) $E_{\rm t}=4.98\times 10^{-3} E_h \,(1092.9\,{\rm cm}^{-1})$, and (d) $E_2=5.0\times 10^{-3} E_h \,(1096.7\,{\rm cm}^{-1})$. 
  }\label{fig:pot_states}
\end{figure}

The following applications are intended to highlight the flexibility of the SimDOC approach when it comes to finding optimal parameters of the control functional itself. This will be demonstrated for final time (Section \ref{sec:tf_opt}) and Hamiltonian (Section \ref{sec:lm_inter_opt}) optimization. Further, we will investigate the influence of the cavity mode frequency, $\omega_{\rm c}$, on the control (Section \ref{sec:omega_c}). In all cases the initial state is given by the vibro--polaritonic ground state, i.e. $|\Phi^{\rm i} \rangle = |0\rangle$. For the target state, a symmetric superposition of the vibro-polaritonic eigenstates $|1\rangle$ and $|2\rangle$ was chosen, i.e. $|\Phi^{\rm t}\rangle = |\Phi_{\rm s}\rangle$, where $|\Phi_{\rm s}\rangle = (|1\rangle + |2\rangle) / \sqrt{2}$. This selection of initial and target states correspond, respectively, to the O-H$\cdots$S and O$\cdots$H-S configuration of TAA. This way we are effectively controlling the H-transfer reaction. The mentioned states are shown for the case of $\eta=0.03$ and  $\omega_{\rm c}=\omega_{10}^q$ in Fig. \ref{fig:pot_states}. Here $\omega_{10}^q=5.76\times 10^{-4}\,E_h \,(126.5\,{\rm cm}^{-1})$ refers to the fundamental transition frequency of the one--dimensional reaction model Hamiltonian from Eq. \eqref{eq.model_transfer_hamiltonian}.

 The vibro-polaritonic target state, $|\Phi^{\rm t}\rangle$, represents a state with a single node along the reaction coordinate, $q$, as observable from Fig. \ref{fig:pot_states}c, but apparently without a cavity mode excitation. However, it is important to notice, that $|\Phi^{\rm t}\rangle$ and all vibro-polaritonic states are displaced along the cavity coordinate, $x_{\rm c}$, as observable from Figs. \ref{fig:pot_states}a-d. This displacement relates to cPES distortion, which leads to a mixing of zeroth-order ($\eta=0$) molecular and cavity eigenstates for $\eta \neq 0$ indicating a non-vanishing photon number expectation value of \textit{virtual} photons, despite the vanishing node along $x_c$.\cite{fischer_saalfrank_2023} In the following simulations, a basis of $N=5$ vibro-polaritonic eigenstates was employed to evaluate matrix elements $\mu_{mn}$  and $\omega_{mn}$ as well as $\varGamma_{m}$. The effective cavity decay rate has been chosen as $\gamma^{-1}=8.39\,$ps yielding a quality factor of $Q=200$ for $\omega_{\rm c}=\omega_{10}^q$.

\begin{figure}[htb!]
\centering
    \includegraphics[width=0.7\linewidth]{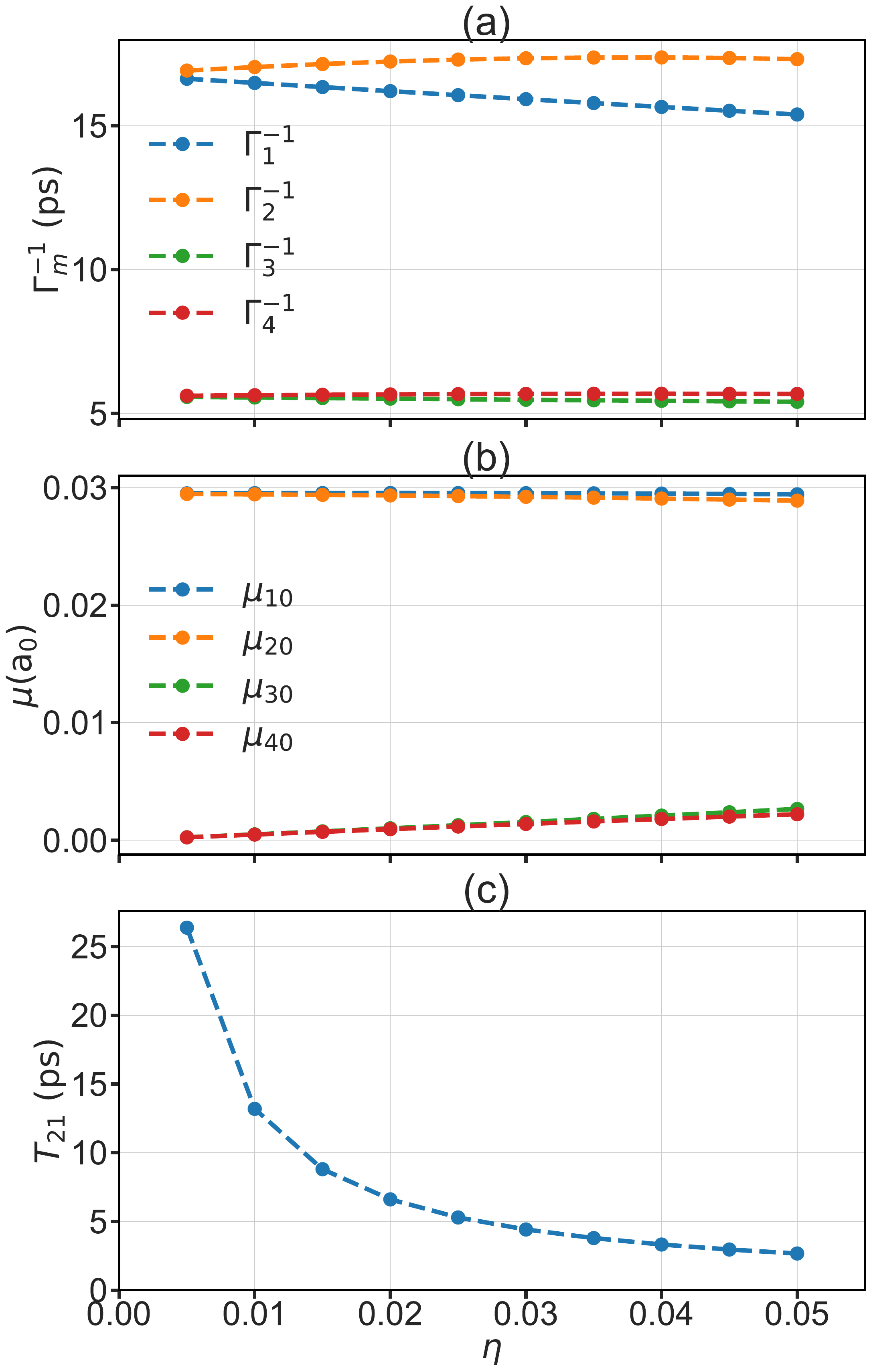}
    \caption{(a) Decay time, (b) dipole moments, and (c) $T_{21}$ for different values of $\eta$ and $\omega_{\rm c}=\omega_{10}^q$.
    } \label{fig:dipole_eta}
\end{figure}

It will be instructive to have at hand the dependence of relevant transition frequencies and dipole moments as well as decay times on the strength of VSC, which are shown in Fig. \ref{fig:dipole_eta} for the case of $\omega_{\rm c}=\omega_{10}^q$. Further, we show in Fig. \ref{fig:dipole_eta}c the time related to the splitting of the vibro-polaritonic states $|1\rangle $ and $|2\rangle $, i.e. $T_{21}=2\pi/\omega_{21}$. Since a superposition of these two states will serve as target of the control, $T_{21}$ defines the relevant system time scale.

\section{Results and Discussion}

\subsection{Final time optimization}\label{sec:tf_opt}

In this section, the capabilities of the  SimDOC approach to 
optimize the final time of the propagation when controlling the state populations is demonstrated. This is motivated by the fact that we included a decay channel for the cavity mode, which provides an additional time scale besides the intrinsic one of the vibro-polaritonic states. Hence the optimal $t_{\rm f}$ should be a compromise between these two times.

As a reference, we computed the total cost for the optimal field as a function of the fixed final time $t_{\rm f}$. We have chosen $\eta=0.03$ and the value of $\omega_{\rm c}=\omega_{10}^q$ has been set to provide  perfect resonance with the fundamental transition frequency  of the reaction Hamiltonian in Eq. \eqref{eq.model_transfer_hamiltonian}.  Respective vibro-polaritonic eigenstates and the target state are shown in Fig.  \ref{fig:pot_states}. Furthermore,  $t_0=0 \,$ps, and $t_{\rm f}$ has been ranged from $2.5\,$ps to $20.6\,$ps in steps of $1.26\,$ps. Notice that for smaller values of $t_{\rm f}$ the total yield approaches zero (not shown) for the given $\kappa$, since  $t_{\rm f}$ would be small compared to the characteristic system time scale.

\begin{figure}[htbp]
\centering
\includegraphics[width=0.65\linewidth]{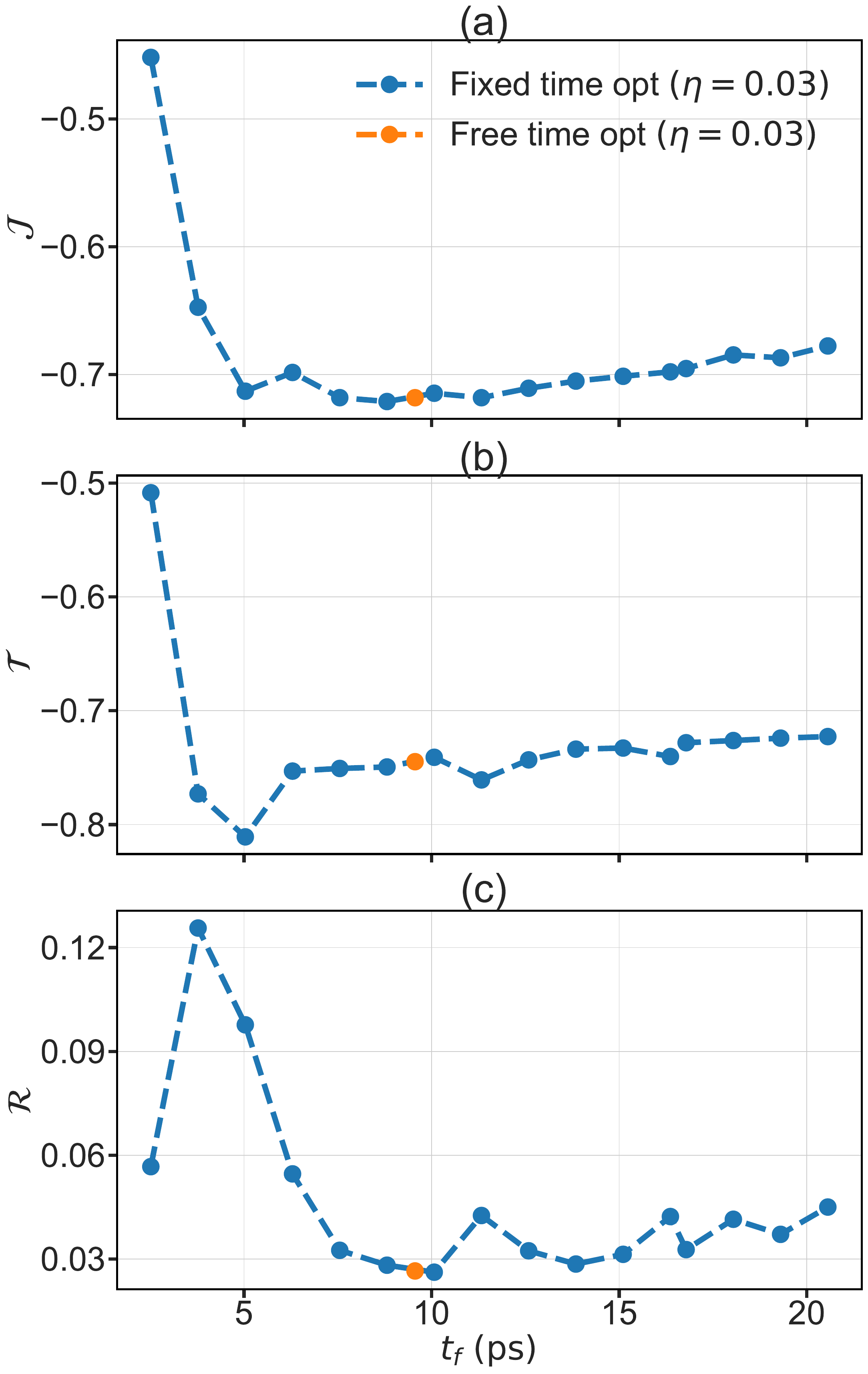}
  \caption{(a) Total cost, (b) terminal cost and (c) running cost as a function of the final time at $\eta=0.03$. For the blue curve, the final time is fixed and not subject to optimization. For the orange dot, the final time is free to vary and it is included in the optimization.}\label{fig:cost_tf}
\end{figure}

The results for the total, terminal, and running  cost as a function of the final time are shown in Fig. \ref{fig:cost_tf}. For this setup, it can be seen that shorter final times (around $2.5\,$ps) lead to higher terminal cost (i.e. smaller target populations, note that the maximum target yield of one corresponds to $\mathcal{T}=-1$) and  higher field intensities, which impact the total cost of the control strategy. For large final times (around $20.6\,$ps), spontaneous decay  due to cavity mirror imperfections starts to dominate. There, one can observe a tendency of  increased terminal cost, while the running cost remains relatively low. In between these two extremes, there is an optimal value for the final time that gives a minimum for the total cost. It should be noted that this general behavior is also expected  for other values of $\eta$ in the range discussed in the next subsection. 

The low target population for small values of $t_{\rm f}$ can be rationalized by inspecting Fig. \ref{fig:pot_states}c. In order to have an efficient population transfer to the target (superposition) state, the field duration should be comparable to the intrinsic time scale given by $T_{21}$. For $\eta=0.03$, we have $T_{21}\sim 5\,$ps and the inverse decay rate $\varGamma_1^{-1} \sim \varGamma_2^{-1} \sim 16 \,$ps. In general, the relationship between these time scales will determine the final population of the target state: The smaller $T_{21}$ relative to  $\varGamma_1^{-1}$ and $\varGamma_2^{-1}$, the less will decay affect the dynamics, such that more population will be accumulated in the target state. This argument holds provided that $t_{\rm f}$ exceeds $T_{21}$. For the small  $t_{\rm f}$ in Fig. \ref{fig:cost_tf}, this is not the case and we observe low target yields (here the decay does not play an important role yet).  A more detailed discussion of the dynamics will be given in the following subsection.

SimDOC enables one to obtain the optimal $t_{\rm f}$ by incorporating the final time in the optimization, i.e. ${\mathcal J}[\bm a, \bm u, \bm k,t_{\rm f}]$, which is a allowed to vary within a given range set by the user. The result of this parameter optimization is  shown as an orange dot in Fig. \ref{fig:cost_tf}. The optimal obtained value of $t_{\rm f}=9.56\,$ps is rather close to the minimum of the ${\mathcal J}(t_{\rm f})$ curve, obtained by fixed $t_{\rm f}$ optimization. This optimal value for $t_{\rm f}$ will depend on the specific value of $\eta$, as shown in Fig. \ref{fig:cost_eta}d below. It is important to notice that with this approach, including the final time in the optimization is as simple as setting the range over which it can vary. There is no need of algorithmic changes as in Ref. \citenum{ndong_time_2014} and this approach comes with no additional computational cost. This flexibility prevents tedious parameter scans. 

%
%
%

\subsection{Cavity-matter interaction strength optimization}\label{sec:lm_inter_opt}

In this section, we examine the impact of the cavity–matter interaction parameter $\eta$ on the optimal control. 
Note that experimentally, $\eta$ can be controlled in certain ranges,  e.g., through the number of molecules in the cavity, the cavity volume or the dielectric constant of the surrounding medium.\cite{ebbesen_2016,fan_quantum_2023}
We have used the same values for $\omega_{\rm c}$, $\kappa$, $\gamma$, $|\Phi^{\rm i} \rangle$ and $|\Phi^{\rm t} \rangle$ as in the previous section. First, fixed $\eta$ optimization will be performed in the range from 0 to 0.05 in steps of 0.005. Afterwards, $\eta$ will be included as a parameter to be optimized. In both cases, the final time $t_{\rm f}$ will also be subject to optimization.

\begin{figure}[thb!]
\centering
    \includegraphics[width=0.5\linewidth]{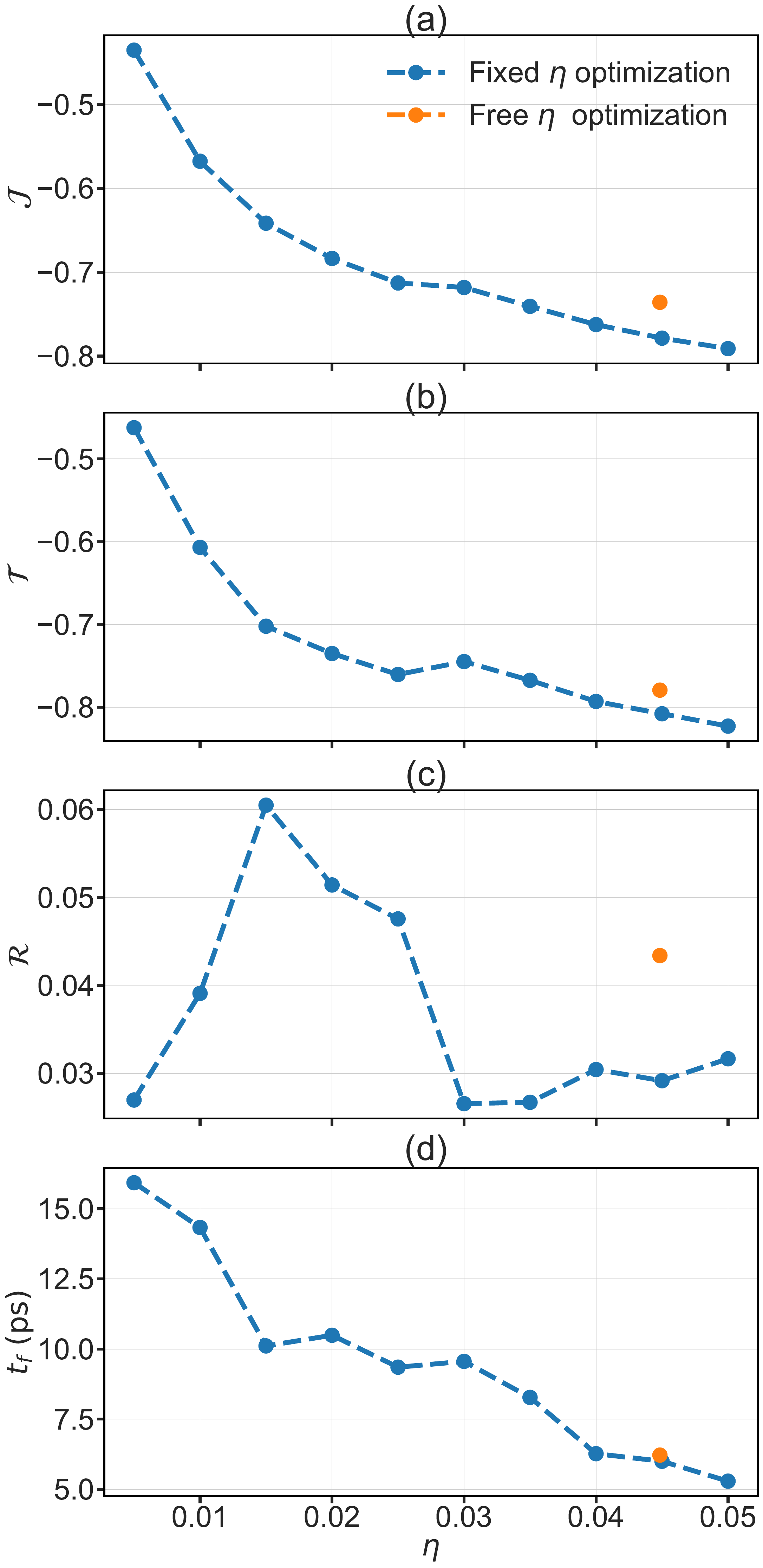}
    \caption{(a) Total cost, (b) terminal cost, (c) running cost and (d) optimal final time as a function of $\eta$, $t_{\rm f}$ is free to be optimized. For the orange dot, $\eta$ has been allowed to vary and it is included in the optimization.} \label{fig:cost_eta}
\end{figure}

\begin{figure}[thb!]
\centering
    \includegraphics[width=0.6\linewidth]{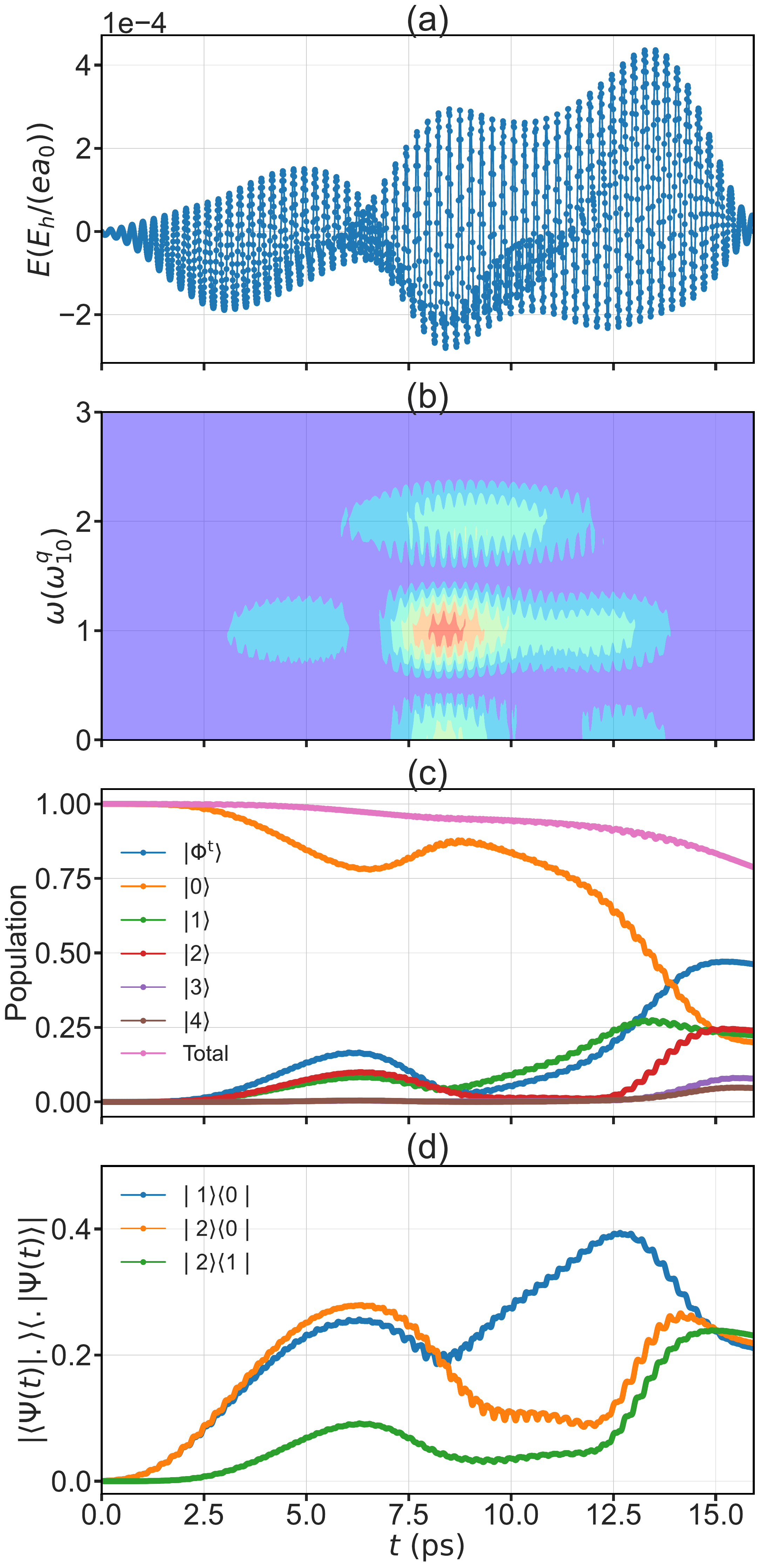}
    \caption{(a) Field, (b) pulse spectrum, (c) population and (d) coherence dynamics for $\eta=0.005$ in Fig.~\ref{fig:cost_eta}.} \label{fig:t_u_pop_0.005}
\end{figure}

 The blue curve in Fig.~\ref{fig:cost_eta}a shows how the total cost changes with $\eta$, provided that the final time is optimized. Overall the total cost decreases with increasing $\eta$. To shed light into this behavior we inspect the dynamics in case of small $\eta=0.005$ given in Fig.~\ref{fig:t_u_pop_0.005}. The optimal field, Fig.~\ref{fig:t_u_pop_0.005}a and b, consists of approximately three subpulses, mostly centered around the transition frequency $\omega_{10}^q$ with some contributions around 0 and $2\,\omega_{10}^q$. According to Fig.~\ref{fig:dipole_eta}, we have $T_{21}\sim 26$~ps, i.e. longer than the decay time of the target state of about 17~ps. Given the discussion of the previous subsection, the relation between these two times is rather unfavorable. The compromise can be seen in Fig.~\ref{fig:t_u_pop_0.005}c and d, where we show population, $|\langle \Psi(t)|m\rangle|^2$, and coherence, $|\langle \Psi(t)|m\rangle\langle  n|\Psi(t)\rangle|$, dynamics, respectively.  Apparently, the effect of decay is reduced by dumping the initially created population of the target state back to the ground state, to re-excite the target state during the final part of the pulse. In passing we note that restricting $t_{\rm f}$ to smaller values during optimization doesn't improve the yield. In fact SimDOC finds a solution where the target state is initially populated and this population is kept almost constant during the simulation interval. Therefore the effect of decay is stronger and thus the overall yield smaller. The respective results are given in Fig.~S1 of the Supplementary Material.

SimDOC allows for including not only $t_{\rm f}$ but also $\eta$ in the optimization procedure, i.e. in the following we choose ${\bm k}=\eta$. Note that in PSOPT, the set of parameters ${\bm k}$ is reserved for parameters that are present in the dynamic equations, as is the case for $\eta$. Since all matrix elements $\mu_{mn}$, energies $E_{m}$ and rates $\varGamma_{m}$ entering the dynamic equations \eqref{eq:re_dyn} and \eqref{eq:im_dyn} depend on $\eta$, they have been computed at each value of $\eta$ presented in Fig.~\ref{fig:cost_eta}. In order to have a continuous dependence of the matrix elements on $\eta$ as needed for optimization, we have used the \textit{smooth\_linear\_interpolation} function provided by PSOPT \cite{becerra_psopt_nodate}, which is the one compatible with the automatic differentiation package ADOLC \cite{griewank_algorithm_1996}, used by the former. In this way, an optimized value of $\eta=\eta_{\rm opt}=0.0448$ with a corresponding value of $t_{\rm f}=6.21\,$ps has been found, which is depicted in Fig.~\ref{fig:cost_eta} as an orange dot.

\begin{figure}[thb!]
\centering
    \includegraphics[width=0.6\linewidth]{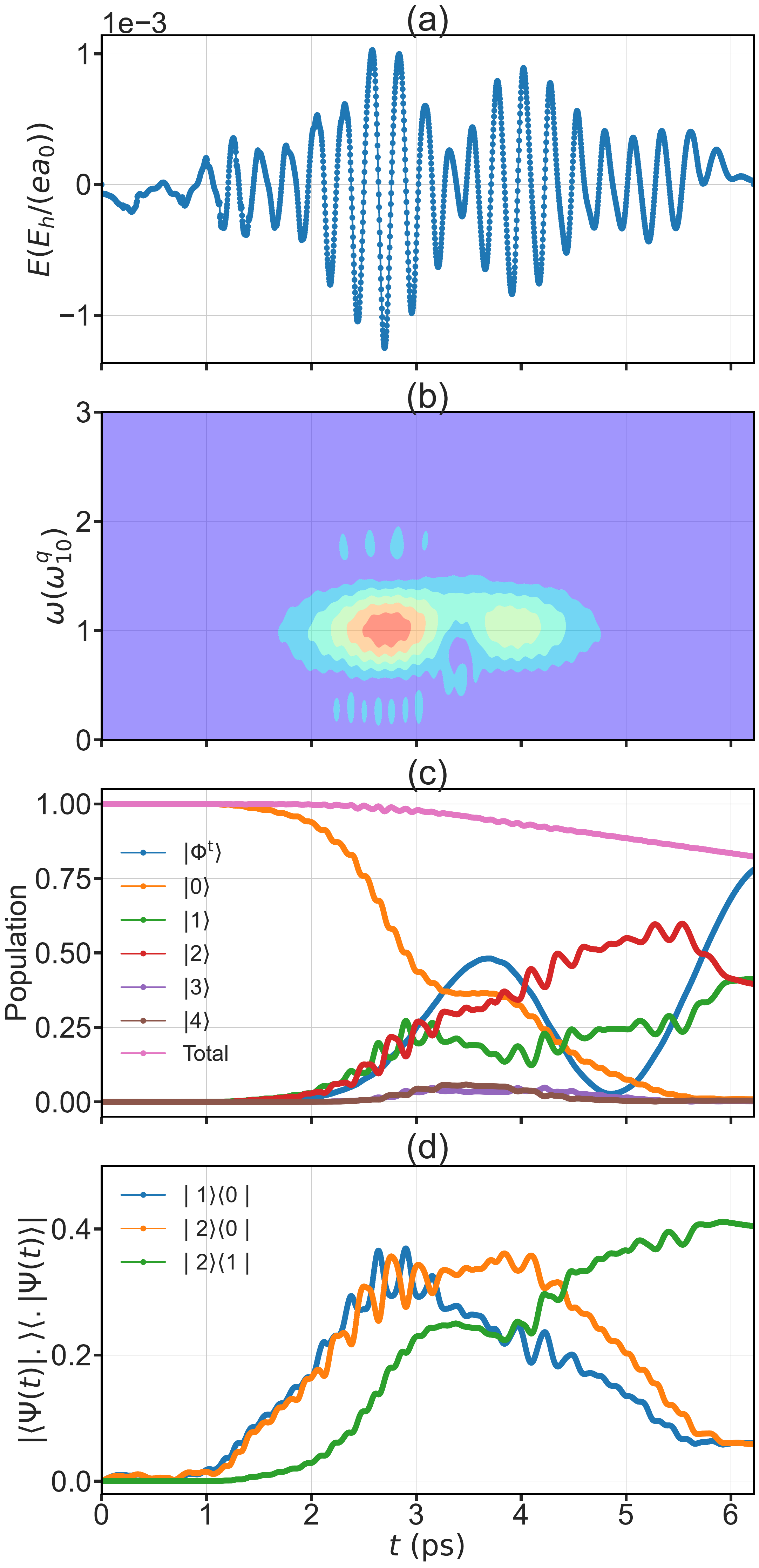}
    \caption{(a) Field, (b) pulse spectrum, (c) population and (d) coherence dynamics for $\eta=\eta_{\rm opt}$ in Fig.~\ref{fig:cost_eta}.} \label{fig:t_u_pop_eta_opt}
\end{figure}

The field, in addition to the time-dependent spectrum as well as selected population and coherence dynamics for $\eta_{\rm opt}$ are shown in Fig.~\ref{fig:t_u_pop_eta_opt}.
The field is essentially composed of two subpulses with the main components around $\omega=\omega_{10}$. For values of $\eta$ around 0.04-0.05 we have for the intrinsic time $T_{21}\sim 3-4$~ps. Hence it is reasonable that the overall pulse duration is shorter compared to the case of small values of $\eta$, in order to reduce the effect of decay. Drawing on the analogy with a $\pi$-pulse the shorter pulse  requires a higher field strength as compared to  Fig. \ref{fig:t_u_pop_0.005}. The population and coherence dynamics in Fig.~\ref{fig:t_u_pop_eta_opt}c and d show that the first pulse creates the proper superposition of the target state already in the middle of the time interval. However, due to the internal dynamics, the superposition changes phase to become $\propto |1\rangle -|2\rangle$ and the target state population is close to zero. Subsequently,  while the pulse still excites the states $|1\rangle$ and $ |2\rangle$ also the phase of the superposition changes to have a yield of about 75\% at the end of the time interval. Note that the final population is limited by the decay, that is, in the present case almost 100~\% of the available population reaches the target state. It is interesting to note that during the second half of the time interval and before reaching the final target state, the population in state $|2\rangle$  is always higher than that of state $|1\rangle$. This can be rationalized by inspecting Fig.~\ref{fig:dipole_eta}a. With increasing $\eta$ the decay times of these two state deviate from each other. The optimization simply prefers to populate the state with the longer lifetime.

From Fig.~\ref{fig:cost_eta} we notice that $\eta_{\rm opt}$ does not correspond to the actual minimum of $\mathcal{J}$ in the given interval. Here it should be noted that the optimization performed with PSOPT is gradient based, i.e. one cannot guarantee that a global minimum is found. The arbitrariness of initial conditions is reduced by sampling of a random ensemble. Still, the control landscape might be rather flat posing a problem for this type of optimization. To make the point in Fig.~S2 of the Supporting Information we show the dynamics for $\eta=0.05$, which gives a smaller value for the total cost in the given $\eta$ interval. The difference in the dynamics between these two values of $\eta$ is rather minor.

\subsection{Influence of the cavity frequency}\label{sec:omega_c}

\begin{figure}[thb!]
\centering
    \includegraphics[width=0.7\linewidth]{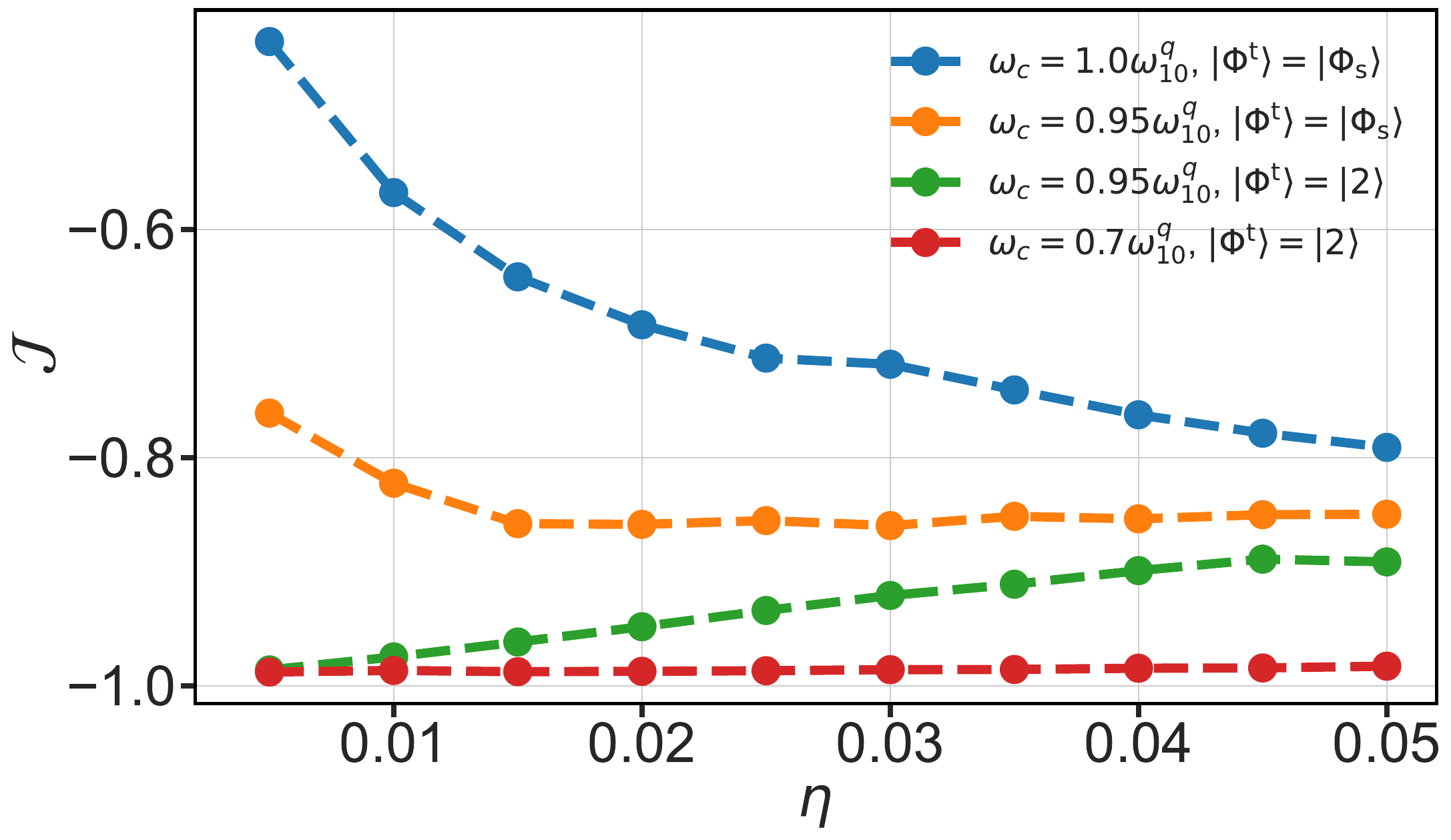}
    \caption{Total cost as a function of  $\eta$ for different values of $\omega_{\rm c}$ as indicated ($|\Phi_{\rm S}\rangle =(|1\rangle + |2\rangle)/\sqrt{2}$).  Target and running cost as well as final time are given in  Fig. S6 in the Suppl. Mat. } \label{fig:total_cost_wc}
\end{figure}

The behavior for the resonant case   $\omega_{\rm c}=\omega_{10}^q$, i.e. the decrease of the total cost $\mathcal J$ with increasing $\eta$ has already been discussed in the previous section. In the following we address the influence of deviations from this strict resonance on the control scenario. 
In general the hybridization between molecular and cavity states depends not only on the coupling strength but also on the detuning, i.e. on the ratio $\eta/|\omega_{\rm c}-\omega_{10}^q|$ (cf. Suppl. Mat. and \citenum{may23}). In the present model, the amount of hybridization determines not only the dipole transition matrix elements and the decay rates, but also the target state state used so far, i.e. $|\Phi^{\rm t}\rangle =(|1\rangle + |2\rangle)/\sqrt{2}$. In case of a large detuning ($\omega_{\rm c}=0.7\omega_{10}^q$ in Fig. \ref{fig:total_cost_wc}) the eigenstates $|1\rangle$ and $|2\rangle$ are of cavity and molecular character, respectively. As a consequence $\mu_{10}\approx 0$ and $\Gamma_2 \approx 0$. In other words the problem is reduced to the molecular one. Using $|\Phi^{\rm t}\rangle = |2\rangle$ the optimization finds a pulse populating the target state by approximately 100\%, independent on the coupling strength $\eta$ (exemplary field and population dynamics are shown in Fig. S4 in the Suppl. Mat.). While this case can be considered to be trivial and not corresponding to the VSC regime, the case of a small detuning is more realistic for the present context. To investigate the effect on the control we have chosen $\omega_{\rm c}=0.95\omega_{10}^q$ in Fig. \ref{fig:total_cost_wc}. Fixing the detuning, upon variation of $\eta$ the extend of hybridization will change. As far as the target state is concerned, to achieve a dominantly molecular state in the product well, it would be appropriate to use $|\Phi^{\rm t}\rangle = |2\rangle$ and $|\Phi^{\rm t}\rangle =(|1\rangle + |2\rangle)/\sqrt{2}$ in the small and large $\eta$ limit, respectively (see also Fig. S3 in the Suppl. Mat.).

The results for these two target states are shown in Fig. \ref{fig:total_cost_wc}. For small $\eta$ the effect of hybridization becomes negligible. Hence the total cost for $|\Phi^{\rm t}\rangle = |2\rangle$ is close to $\mathcal J =-1$. In case of $|\Phi^{\rm t}\rangle =(|1\rangle + |2\rangle)/\sqrt{2}$ the state $|1\rangle$ cannot be effectively excited and thus  $\mathcal J$ increases. For large $\eta$ and the given small detuning, hybridization becomes dominant. Here the curve for $|\Phi^{\rm t}\rangle =(|1\rangle + |2\rangle)/\sqrt{2}$ approaches the resonant case (exemplary field and population dynamics are shown in Fig. S5 in the Suppl. Mat.). On the other hand, for $|\Phi^{\rm t}\rangle = |2\rangle$ the reaction yield decreases as decay becomes more effective.

\section{Conclusions}
\label{sec:conlcusions}
We have studied the external field-driven intramolecular hydrogen transfer dynamics in an asymmetric potential under vibrational strong coupling with a single cavity mode, initially discussed in Ref. \citenum{fischer_saalfrank_2023}, by means of the SimDOC approach. It was shown that, in principle, control utilizing the hybridization between molecular and cavity states is possible, but limited by the presence of a decay channel for the cavity mode. While this appears to be not surprising, the actual focus has been on demonstrating the power of the SimDOC approach itself. 

For example, choosing an appropriate final time when controlling a system in the presence of different system time scales (here eigenstate dynamics and  decay) is key for achieving an optimal control. Using the SimDOC approach there is no need to scan the final time to obtain the optimal one, but it is calculated directly within the optimization procedure. Importantly, this generates no overhead with respect calculation time.

Furthermore, the SimDOC approach has been proven useful when implementing the optimization of certain parameters within the model. This was demonstrated for the  case of  the dimensionless light–matter interaction parameter. In other words, the VSC Hamiltonian offers a means to shape the control landscape by changing the coupling strength. However, in general, any other system parameter can be selected as long as a smooth interpolation on the desired parameter for the Hamiltonian matrix is possible (here $\mu_{mn}$, $E_{m}$ and rates $\varGamma_{m}$). 

It should be noted, however, that SimDOC is a gradient-based method. This implies that a local minimum will be found during optimization. Since the present implementation involves scanning random initial conditions for optimization, the danger of being trapped in a less than optimal solution is reduced. 

As far as the external field control of molecules in cavities under VSC is concerned, there are several directions to improve the present model. Most notably, the relaxation model could be modified by changing from a wavefunction to a density matrix description. Since the actual type of dynamics' equations is not restricted in SimDOC, this extension would be straightforward.

\section*{Acknowledgments}
We acknowledge support by Deutsche Forschungsgemeinschaft (DFG) through Sonderforschungsbereich 1636, project A05, 'Understanding and controlling reactivity under vibrational and electronic strong coupling' (P.S.) and the project Ku952/10-1 (O.K.).


\end{document}